\documentclass[%
 amsmath,amssymb,
 aip,
]{revtex4-2}

\usepackage{graphicx}
\usepackage{dcolumn}
\usepackage{bm}

\usepackage[utf8]{inputenc}
\usepackage[T1]{fontenc}
\usepackage{mathptmx}
\usepackage{comment}
\usepackage{color}
\usepackage[dvipsnames]{xcolor}

\begin{document}

\title{Elucidation of Relaxation Dynamics Beyond Equilibrium Through AI-informed X-ray Photon Correlation Spectroscopy}

\author{James P. Horwath}%
\affiliation{ 
Advanced Photon Source, Argonne National Laboratory, Lemont, IL 60439.
}
\author{Xiao-Min Lin}%
\affiliation{ 
Center for Nanoscale Materials, Argonne National Laboratory, Lemont, IL 60439.
}
\author{Hongrui He}
\affiliation{Materials Science Division and Center for Molecular Engineering, Argonne National Laboratory, Lemont, IL 60439.}
\affiliation{Pritzker School of Molecular Engineering, University of Chicago, Chicago, IL 60637.}

\author{Qingteng Zhang}%
\affiliation{ 
Advanced Photon Source, Argonne National Laboratory, Lemont, IL 60439.
}
\author{Eric M. Dufresne}%
\affiliation{ 
Advanced Photon Source, Argonne National Laboratory, Lemont, IL 60439.
}

\author{Miaoqi Chu}
\affiliation{
Advanced Photon Source, Argonne National Laboratory, Lemont, IL 60439.
}

\author{Subramanian K.R.S. Sankaranarayanan}%
\affiliation{ 
Center for Nanoscale Materials, Argonne National Laboratory, Lemont, IL 60439.
}
\affiliation{ 
Department of Mechanical and Industrial Engineering, University of Illinois, Chicago, IL 60607
}
\author{Wei Chen}
\affiliation{Materials Science Division and Center for Molecular Engineering, Argonne National Laboratory, Lemont, IL 60439.}
\affiliation{Pritzker School of Molecular Engineering, University of Chicago, Chicago, IL 60637.}

\author{Suresh Narayanan}
\affiliation{ 
Advanced Photon Source, Argonne National Laboratory, Lemont, IL 60439.
}%
\author{Mathew J. Cherukara}
\email{mcherukara@anl.gov, sureshn@anl.gov, jhorwath@anl.gov}
\affiliation{ 
Advanced Photon Source, Argonne National Laboratory, Lemont, IL 60439.
}%


\date{\today}

\begin{abstract}
Understanding and interpreting dynamics of functional materials \textit{in situ} is a grand challenge in physics and materials science due to the difficulty of experimentally probing materials at varied length and time scales.  X-ray photon correlation spectroscopy (XPCS) is uniquely well-suited for characterizing materials dynamics over wide-ranging time scales, however spatial and temporal heterogeneity in material behavior can make interpretation of experimental XPCS data difficult.  In this work we have developed an unsupervised deep learning (DL) framework for automated classification and interpretation of relaxation dynamics from experimental data without requiring any prior physical knowledge of the system behavior. We demonstrate how this method can be used to rapidly explore large datasets to identify samples of interest, and we apply this approach to directly correlate bulk properties of a model system to microscopic dynamics.  Importantly, this DL framework is material and process agnostic, marking a concrete step towards autonomous materials discovery.
\end{abstract}

\maketitle
\section{\label{sec:intro}Introduction} 
Structure-property relationships are the core of materials science and condensed matter physics, however, defects and disorder make it difficult to describe real materials with simple analytical models.
This challenge is even more prominent in metastable and out-of-equilibrium materials, where prevailing theoretical frameworks cannot be used to accurately model the system dynamics. A prime example of this is studying relaxation in complex non-newtonian fluids. Though such materials have wide-ranging industrial applications, from technical coatings to food preparation, their rheological properties and relaxation behavior remain poorly understood\cite{joshi_dynamics_2014,wen_dynamics_2015}.  While standard rheology methods and x-ray scattering experiments give dependable characterization of the bulk properties and microstructure, respectively, the stochastic nature of microstructrual changes which lead to unpredictable, non-monotonic changes in viscosity makes it difficult to correlate microstructural fluctuations with macroscopic property measurements\cite{besseling_oscillatory_2012,xu_relation_2013,lee_unraveling_2018}.
Defining and explaining the link between microstructure and measurable bulk responses is an important, interesting, and challenging problem from both science and engineering perspectives\cite{manoharan_colloidal_2015,abou_aging_2001,philippe_glass_2018}.

X-ray photon correlation spectroscopy (XPCS), a coherent x-ray scattering technique built on the same fundamental mechanisms as Dynamic light scattering (DLS), measures complex fluid dynamics via temporal decorrelation of scattered x-ray intensities\cite{chu_dynamic_2008,goldburg_dynamic_1999,shpyrko_x-ray_2014}.  XPCS measurements span a spatio-temporal range from sub-nm $\sim \mu$m and $\mu$s $\sim$ hours. It can further be combined with sophisticated \textit{in-situ} environments thanks to the use of hard x-rays with high penetration power\cite{Sheyfer2020-yb,Girelli2021-zz,Lehmkuhler2020-ir,Dallari2021-ay,zhang_dynamics_2018,lee_unraveling_2018,Ju2019-xc,Myint2021-mt}. 

While XPCS is ideally suited to studying relaxation in complex fluids due to its compatibility with \textit{in situ} rheometry, mesoscale spatial resolution, and ability to capture structure transition with high temporal resolution over very long time scales, analysis of XPCS data can be difficult.  This challenge is rooted in the experimental limits of XPCS characterization, and the spatial heterogeneity of local structure and dynamics.  While dynamics in a meta-stable or non-equilibrium systems  vary spatially within in a sample, XPCS experiments only probe local environments within a field of view limited to the size of the coherent x-ray beam (typically a few microns).  Under steady-state conditions, the initial observation time does not matter since dynamics are assumed to be constant through time.  Additionally, since the x-ray probe is smaller than the specimen being studied, our analysis relies on the assumption that dynamics are also spatially homogeneous. Under these assumptions, traditional XPCS analysis is based upon calculating the mean correlation intensity ($g_2$) at all equivalent delay times ($\tau = t_1 - t_2$), to produce a one-dimensional plot of $g_2$ vs $\tau$.  From here, analytical models describing the dynamic decorrelation can be assumed, and experimental data is fit to the model to extract physical parameters.  However, the dynamics of relaxation are often strongly outside of equilibrium conditions, and therefore other analysis methods are required.  For non-equilibrium XPCS analysis, the pair-wise correlation is calculated across a time sequence of scattering frames and displayed in a two-dimensional figure called a two-time correlation ($C_2(t_1,t_2)$).  As $C_2$ are capable of describing any type of relaxation dynamics, they can provide a dynamic "fingerprint" of the non-equilibrium system at any given experimental time, and a variety of analyses are being considered to take advantage of the information-rich second-order correlations which describe changes between specific time points\cite{bikondoa_use_2017,Zhang2017-oz, Ruta2020-oa, Dallari2020-sf,hu_cross-correlation_2021,Perakis2017-lt,Dallari2020-sf,Song2022-vy}. Still, the amount of human adjudication required for interpretation of results from such advanced XPCS analysis methods, as well as the amount of data collected in synchrotron experiments, presents a significant barrier to the development of a more quantitative physical understanding of dynamics in complex fluids; without the ability to observe structure and dynamics at many points in the sample, it is difficult to link microstructural changes to bulk properties which represent the average across the entire system.  To further complicate the matter, the variety of patterns shown in experimental $C_2$ of our model system vary drastically such that even visual identification of relationships between data points is difficult (see Figure \ref{fig:schematic} C for a sample of $C_2$ data).  The limitation imposed by data interpretation bandwidth will become even more pronounced with the use of high-frame-rate, large-pixel-array x-ray detectors and the world-wide commissioning of ultra-brilliant fourth-generation undulator x-ray sources \cite{Zinn2018-vh,Pennicard2018-qn,Nakaye2021-oc,Leonarski2018-ya,Dooling2022-jb,Schroer2018-ty,Martensson2018-kt,Chenevier2018-et}.

Recent years have seen a tremendous increase in the application of machine learning (ML) methods to scientific data with applications ranging from assisting medical diagnosis and guiding autonomous vehicles, to solving fundamental physical problems\cite{kumar_artificial_2022,ma_artificial_2020,cranmer_discovering_2020}. Specific to x-ray characterization, ML methods are being used across nearly every characterization technique\cite{benmore_advancing_2022}.  Examples include the use of ML to determine the structure-property relationship \cite{decost2017computer,Schmidt2019,Meredig2019, wang2020machine,ma2020accelerated}, to accelerate and enhance coherent characterization techniques\cite{yao2022autophasenn, cherukara_ai-enabled_2020,Henry_chan_DL,Zhou2021,cherukara2018real,wu2021three}, accelerate emission spectroscopy, reduce dose and noise in tomography and accelerate Bragg peak fitting\cite{hwang_axeap_2022,yang_low-dose_2018,liu_tomogan_2020,liu_braggnn_2022}. Recent work has demonstrated the use of ML to denoise $C_2$, which lead to significant improvement on the quantitative interpretation of the XPCS results and detection of anomalous results \cite{konstantinova_noise_2021,konstantinova_machine_2022}. 

Here, we develop an automated, unsupervised ML workflow for the automated classification of experimental \textit{rheo}-XPCS datasets.  We demonstrate the development of a convolutional autoencoder (AE) for encoding $C_2$ into a reduced space. We then apply K-Means Clustering to classify datapoints based on their position in the later space.  Next, we illustrate the utility of this type of analysis in a representative use case, namely non-equilibrium dynamics. By classifying $C_2$ and comparing transitions between classes as a function of time to rheological measurements showing the evolution of shear stress within the material we show the correlation between jammed structures and relaxation rates.  Finally, we show how our method can be used to take in user-specified $C_2$ of interest, and return other samples from the dataset in order of similarity; from here, relationships between structure and rheology can be inferred by observing distributions of experimental parameters within, and across, groups of similar images.

\section{\label{sec:results}Results}

\subsection{\label{sec:rheo-xpcs} Relaxation of colloidal glass probed with \textit{in-situ} XPCS}
The experimental setup consists of an instrument that combines a rheometer and XPCS measurement simultaneously, i.e. \textit{rheo}-XPCS, to study the structural response of a colloidal glassy system under shear.  In these experiments, a model colloidal glass made up of silica spheres suspended in polyethylene glycol is loaded into a Couette shear cell.  A rheometer drives the shear cell under various strain rate-controlled protocols, and measures the shear stress response and viscosity as a function of time.  After applying the shear protocol the shear rate is set to zero, and x-ray scattering data is collected simultaneously with shear stress measurement to observe both the mechanical and structural relaxation processes (Figure \ref{fig:schematic} A).   During a XPCS experiment, a "movie" of scattered x-ray intensity is acquired in a continuous time sequence using a pixelated photon-counting x-ray detector. The scattered intensity exhibits a "speckled" optical interference pattern due to the coherence of the beam \cite{brown_speckle_1997}.  The correlation between collected frames is calculated based on the pixel-wise intensity at a given scattering wavevector. Dynamics can then be determined based on the lifetime of the correlation\cite{leheny_xpcs_2012}.  In a two-time correlation function ($C_2$), the correlation is calculated between all possible pairs of frames, and displays data as a two-dimensional image, where experimental time is represented on both axes, the diagonal corresponds to the auto-correlation of one frame with itself (producing high intensity along the diagonal, as seen in Fig.~\ref{fig:schematic}(B). Details regarding the XPCS measurement and the correlation algorithms can be found in Section~\ref{sec:xpcs_measurement} and in previous literature \cite{Zhang2017-oz,Khan2018-aw}. 

Dynamic heterogeneity and non-linear rheological response, both resulting from highly heterogeneous distributions of constituent particles and their local motion, are well known in glassy systems\cite{berthier_dynamical_2011}.  Avalanches in hard sphere glasses can be triggered by motion of particles in a very localized area, which subsequently influences particles in other areas\cite{sanz_avalanches_2014}.  There are very few experimental techniques that can study this heterogeneity, among which XPCS has a definite advantage of providing information with high spatial and temporal resolution\cite{hoshino_dynamical_2020}. In addition to the experimental challenges of building a full description of the relationship between microscopic dynamics and observable bulk properties, the wide variety of dynamics captured in XPCS data (Figure \ref{fig:schematic}C), and their wide range of appearances, makes analysis of available data extremely complicated.

In the following sections, we describe the use of unsupervised Machine Learning to process collections of experimental \textit{rheo}-XPCS data, with the goal of automating the arduous and expert-driven process of interpreting and classifying the wide distribution of $C_2$ topologies which come from a single experiment.

\subsection{\label{sec:optimization}Unsupervised Deep Learning to elucidate relaxation dynamics}

Machine learning models generally can fit into either the supervised, or unsupervised learning paradigms.  In supervised learning, scientists provide a labeled dataset which is used to optimize the model weights based on the difference between model predictions and the provided \textit{ground truth}.  Unsupervised learning is used in cases where labeled data is unavailable or difficult to produce, and algorithms generally aim to distill features of the raw data, identify statistical trends across the dataset, or cluster the dataset based on the relatedness between data points.  Unsupervised learning presents opportunities for reaping the pattern recognition and processing acceleration benefits of machine learning without requiring labeled data or even physical understanding of the system\cite{wang_discovering_2016,schmarje_survey_2021}. This is incredibly useful for understanding structural dynamics from experimental $C_2$, since physical interpretation of non-equilibrium $C_2$ is difficult and the associated dynamics are poorly understood.  For experimental data that can be represented as images, such as XPCS $C_2$, convolutional neural networks (CNN) are able to accurately encode spatial information, and take advantage of the expressive power of deep learning to provide accurate and adaptive understanding of scientific data\cite{long_fully_nodate}.  

\begin{figure}[h]
    \centering
    \includegraphics[width=\textwidth]{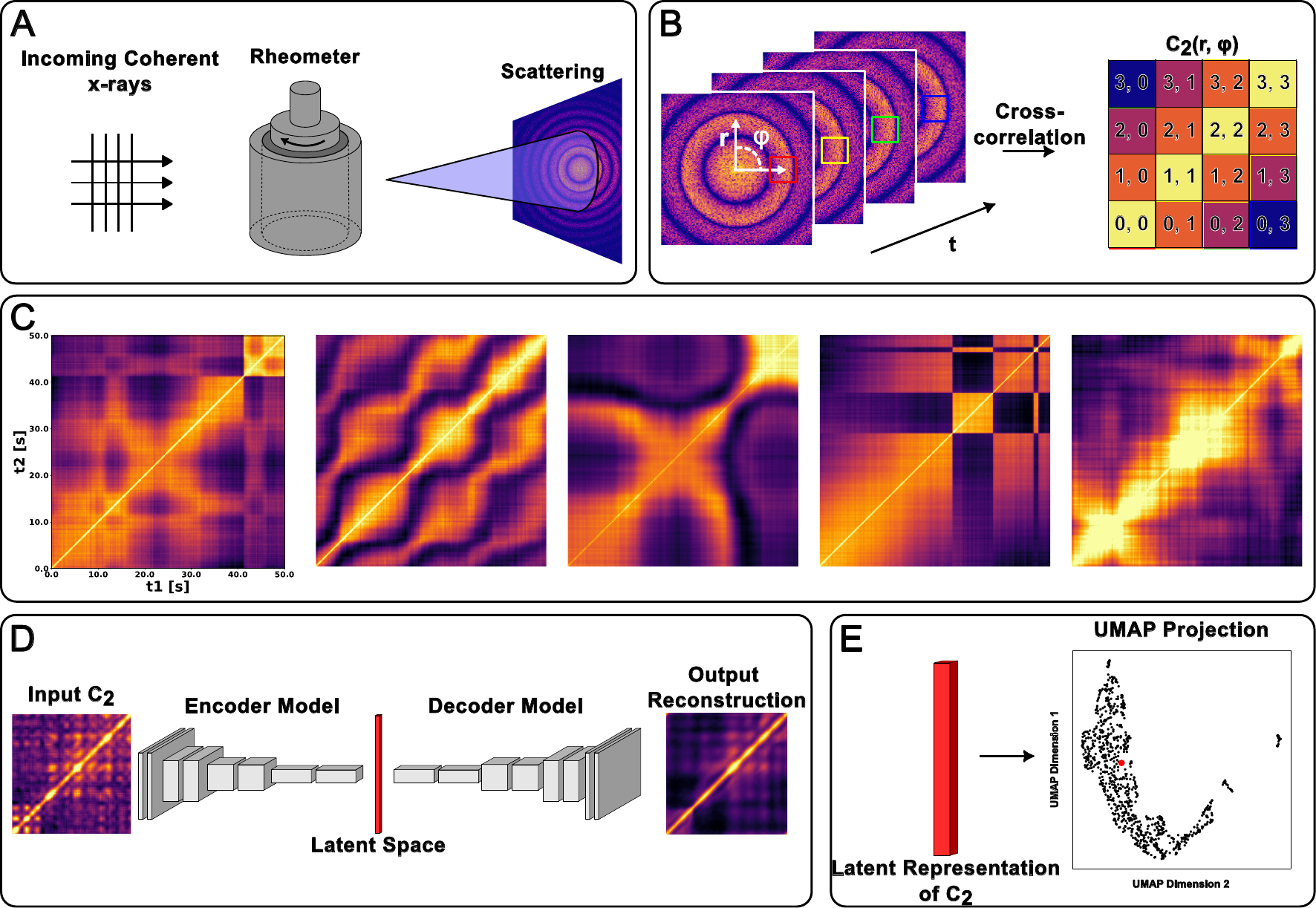}
    \caption{Schematic of the experimental setup and machine learning workflow.  A.) In \textit{rheo}-XPCS, a rheometer is placed in the beam path so that coherent x-rays scatter off the relaxing sample.  B.) XPCS two-time correlations are calculated by correlating intensity in a specific scattering region over time. Correlations are mapped as $C_2$.  C.) shows a sample of experimental $C_2$ to illustrate the wide variation in dynamics seen in non-equilibrium XPCS. The time scale bar in the left-most $C_2$  applies for all other images.  D.) The autoencoder is trained to reproduce raw $C_2$, and the learned latent representation is used to cluster and classify datapoints (E.).}
    \label{fig:schematic}
\end{figure}

Our unsupervised ML workflow uses a convolutional autoencoder to generate a feature-rich latent space representation from which we perform further analysis. Convolutional autoencoders are well suited to our task of automated pattern recognition with no prior information.  These architectures consist of an encoder model, which uses a series of convolutional layers to encode raw image data into a feature-rich latent representation, and decoder model which takes the learned latent features as input and attempts to reconstruct the image based on this reduced representation; loss is calculated based on the difference between the original input data and the output reconstruction.  While autoencoders have proven successful in a variety of computer vision tasks, they are also used as flexible compression or dimensionality-reduction algorithms due to their ability to extract and encode important image features\cite{chen_deep_2021,cheng_deep_2018,gondara_medical_2016}. In these cases of compression and automated classification, the final output image is not considered outside of the training process and all analysis is performed on the latent space of the trained model based on the understanding that the latent space of a well-trained model expresses the entire distribution of the training data\cite{yeh_learning_nodate,patel_latent_2013}.  We adapt this approach to encode experimental $C_2$, and classify data based on their latent representation.  A schematic of our autoencoder and latent space analysis is provided in Figure~\ref{fig:schematic} D and E.

 We have employed an hourglass-style convolutional autoencoder which compresses 256 x 256 pixel $C_2$ into a 64-dimensional latent space, and then decodes the latent representation to reproduce the input data. Further details of model optimization and data augmentation are presented in Section \ref{sec:CNN_development}. The sample reconstruction shown in Figure \ref{fig:schematic}D appears as a denoised version of the raw input. More example of experimental $C_2$ and corresponding AE outputs can be seen in Supplemental Figure 1.  These result signify accurate model performance - random noise fluctuations in an image will be difficult to capture in image filters optimized for performance on an entire dataset, so the absence of noise suggests that learned filters focus on more important image features. While imperfect, the output reconstruction from our optimized architecture maintained long-range features and time-scale information such as the position of changes in the width of the diagonal correlation band, and off-diagonal patterns.

\subsection{$C_2$ Classification and Latent Space Analysis}
After training the optimized model, a new dataset, corresponding to Rheo-XPCS measurement from a single rheological shear cycle was fed through the encoder model.  Unsupervised classification of the new $C_2$ dataset was performed by applying k-means clustering algorithm directly to the latent representation of the data, determining the ideal number of clusters using the elbow method (Supplemental Figure 3).  This showed that four - six clusters was the ideal number, with fewer clusters separating data solely based on image intensity and more clusters separating the data into unrealistically small groupings.

\begin{figure}[h]
    \centering
    \includegraphics[width=\textwidth]{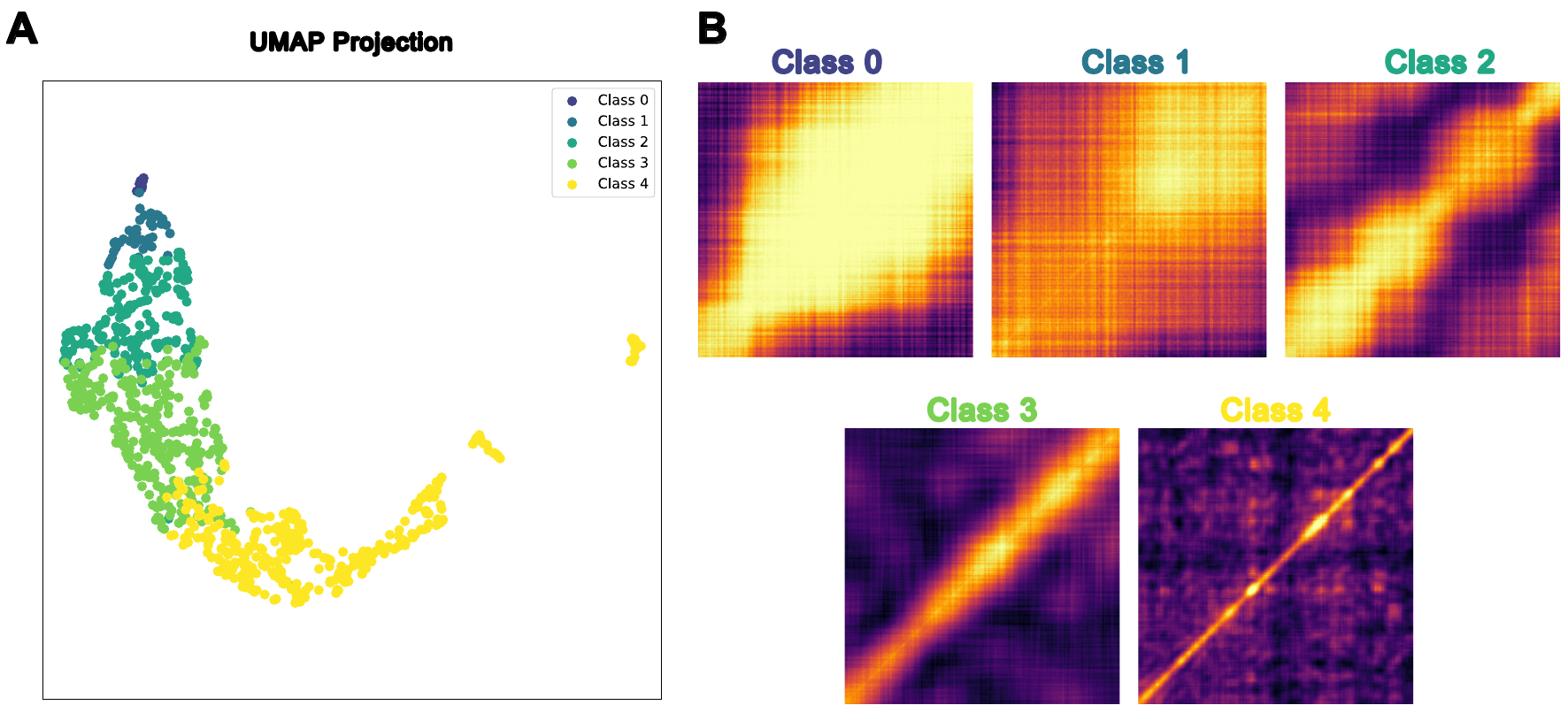}
    \caption{A.) shows the UMAP visualization of the latent space.  Point colors correspond to the clusters labels, and match the image labels in B.).  Each image in B.) shows a random sample from each class.}
    \label{fig:initial_clustering}
\end{figure}

Due to the high dimensionality of even the latent representation, further embedding is required to visualize the distribution of the encoded data and the clustering results.  We used Uniform Manifold Approximation and Projection (UMAP) to transform the latent space of the dataset into two dimensions; this visualization is shown in Figure \ref{fig:initial_clustering}\cite{mcinnes_umap_2020}.
UMAP is closely related to t-distributed stochastic Neighbor Embedding (tSNE), a more common method for non-linear dimensionality reduction\cite{van_der_maaten_visualizing_2008}.  Both of these methods consider the local structure of the data distribution and attempt to project data points onto a lower dimensional manifold, however, in comparison to tSNE, UMAP distorts the data distribution such that it is uniformly distributed in the projection space.  This helps maintain the global structure of the dataset, and generates projections which are more stable against variation in initialization and hyperparameters than those generated by tSNE\cite{nguyen_symmetry-aware_2021}.  This visualization allows us to qualitatively check the accuracy of the clustering results by seeing whether optimal cluster centers coincide with the densest regions of the UMAP embedding, however, it is important to note that distances in these embedding spaces cannot be quantitatively compared, and only serve a qualitative metric of similarity between data points\cite{wattenberg_how_2016}.

Viewing images from each class (Figure \ref{fig:initial_clustering}) shows that relaxation times decrease with increasing cluster label.  Following the trajectory across the UMAP distribution in Figure \ref{fig:initial_clustering}A from left to right, we can see the transition between nearly stationary dynamics in Class 0 (high correlation relates to slow structural changes) to slow evolution in Class 1 (seen as flat $C_2$ features with lower intensity), to increasingly fast evolution in Classes 2, 3, and 4.  Further, the nearly-continuous transition between clusters illustrates that the groupings defined by the k-means algorithm contain physically similar data, and maintain physical meaning between the clusters.

\subsection{Probing non-equilibrium dynamics using trained ML model}
With a trained autoencoder and the ability to rapidly encode and classify experimental data in hand, we now describe how this approach can be used to understand dynamics and explore patterns in large datasets.

\begin{figure}[h]
    \centering
    \includegraphics[width=\textwidth]{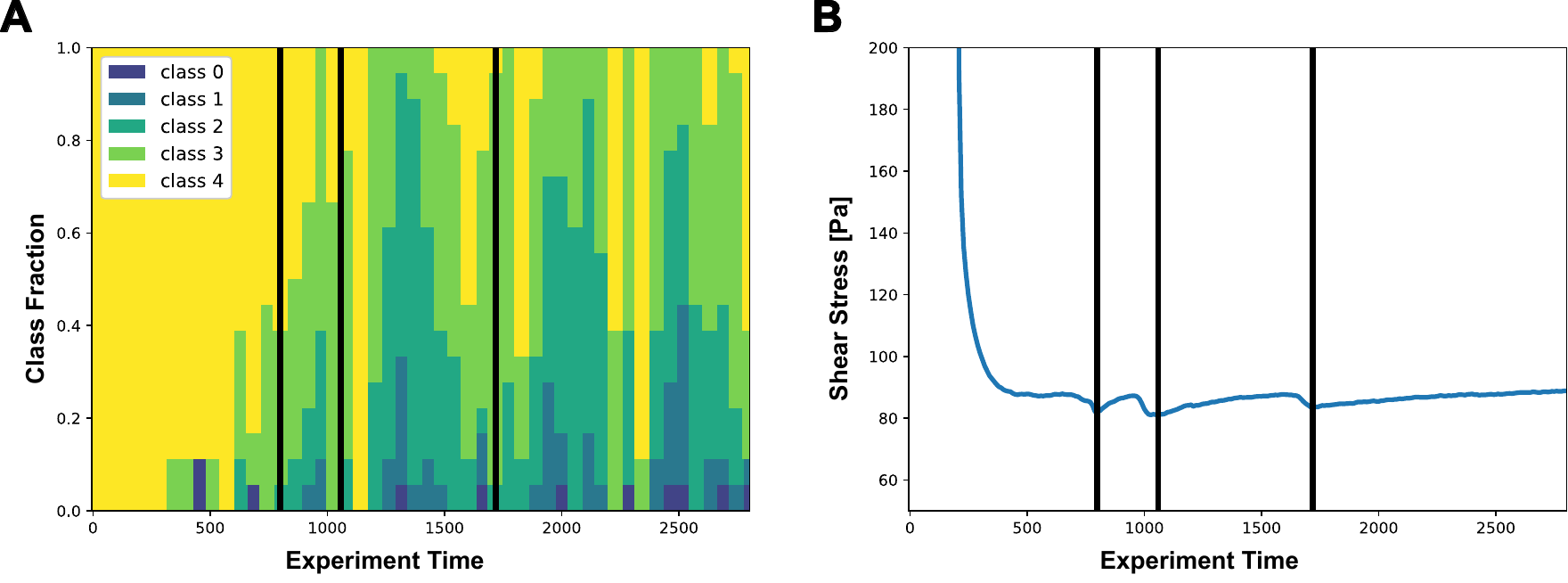}
    \caption{Analyzing the distribution of $C_2$ classes as a function of time (A), alongside the shear stress measured in relaxing fluid (B) allows us to relate changes in relaxation dynamics to fluctuations in the mechanical response.  Colors/classes correspond to those in Figure \ref{fig:initial_clustering}.  Vertical black lines in both plots corresponds to times with shear stress minima. Peaks of green between shear stress minima represent slow dynamics as shear stress builds up.}
    \label{fig:rheo_analysis}
\end{figure}

Bringing us one step closer to our goal of bridging information across measurement modalities and length scales, our first test aims to understand how fluctuations in rheological measurements correspond with the evolution of the structure and local dynamics of the suspension.  After a steady shear into the shear thickening regime, we set the shear rate to zero, and monitor the decay of shear stress in the system.  Typically, the relaxation of shear stress shows a non-monotonic behavior (Figure \ref{fig:rheo_analysis}). One may suspect that sudden changes in the relaxation rate may be a result of changes in microstructure and local dynamics which either enhance or inhibit the release of shear stress. Considering that XPCS and rheology data are collected simultaneously, we can directly compare dynamics and rheological response through time.  Moreover, by extracting unique $C_2$ from varying scattering vectors during a single timestep (described in Section \ref{sec:xpcs_measurement}), we can use our unsupervised clustering method to classify the dynamics from the collection of $C_2$ capture simultaneously the distribution of relaxation behaviors as a function of time.  The percentage of each class at each time step is shown as a vertical, color-coded bar in Figure \ref{fig:rheo_analysis} A.  Visualization of samples from each class (Figure \ref{fig:initial_clustering}) shows that classes correspond to different dynamic behavior, with relaxation rate increasing as class number increases (nominally, relaxation rate is related to the time scale associated with decay of the correlation fucntion perpendicular to the main diagonal). Observing changes in the class distribution in Figure \ref{fig:rheo_analysis} we see fast dynamics in the first stage where shear stress rapidly decreases, which suggests that the structure is rapidly evolving to accommodate and release stress in the system.  In the stable stress region (t > 500 s), there are clustering of slow dynamics regions (peaks of green and dark green) interrupted by a short period of fast dynamics (yellow). This indicates that in the shear-induced glassy state, local intermittent particle rearrangement is happening inside the sample.  Additionally, our clustering analysis shows that rheological shear stress fluctuations (black vertical lines in Figure \ref{fig:rheo_analysis}A) coincide with regions of transition between slow and fast dynamics.  The correspondence between slow dynamics and sudden increase in the shear stress may indicate the formation of jammed regions, and explain why many samples show finite yield stress after relaxation. While our clustering results do not perfectly describe the rheological response (see green peaks near t = 2000 and t = 2400), the mismatch can be attributed to the fact that XPCS tracks small-scale local changes which may have a small impact on the global material properties. While previous research has used \textit{rheo}-XPCS to study the link between dynamics, internal stress, and jammed structure, to our knowledge this is the first example of using \textit{in situ} x-ray scattering to relate microscopic dynamics with macroscopic response without requiring the assumption of equilibrium or phenomenological physical models to quantify dynamics\cite{chen_microscopic_2020,rogers_microscopic_2018}.

\begin{figure}[h]
    \centering
    \includegraphics[width=\textwidth]{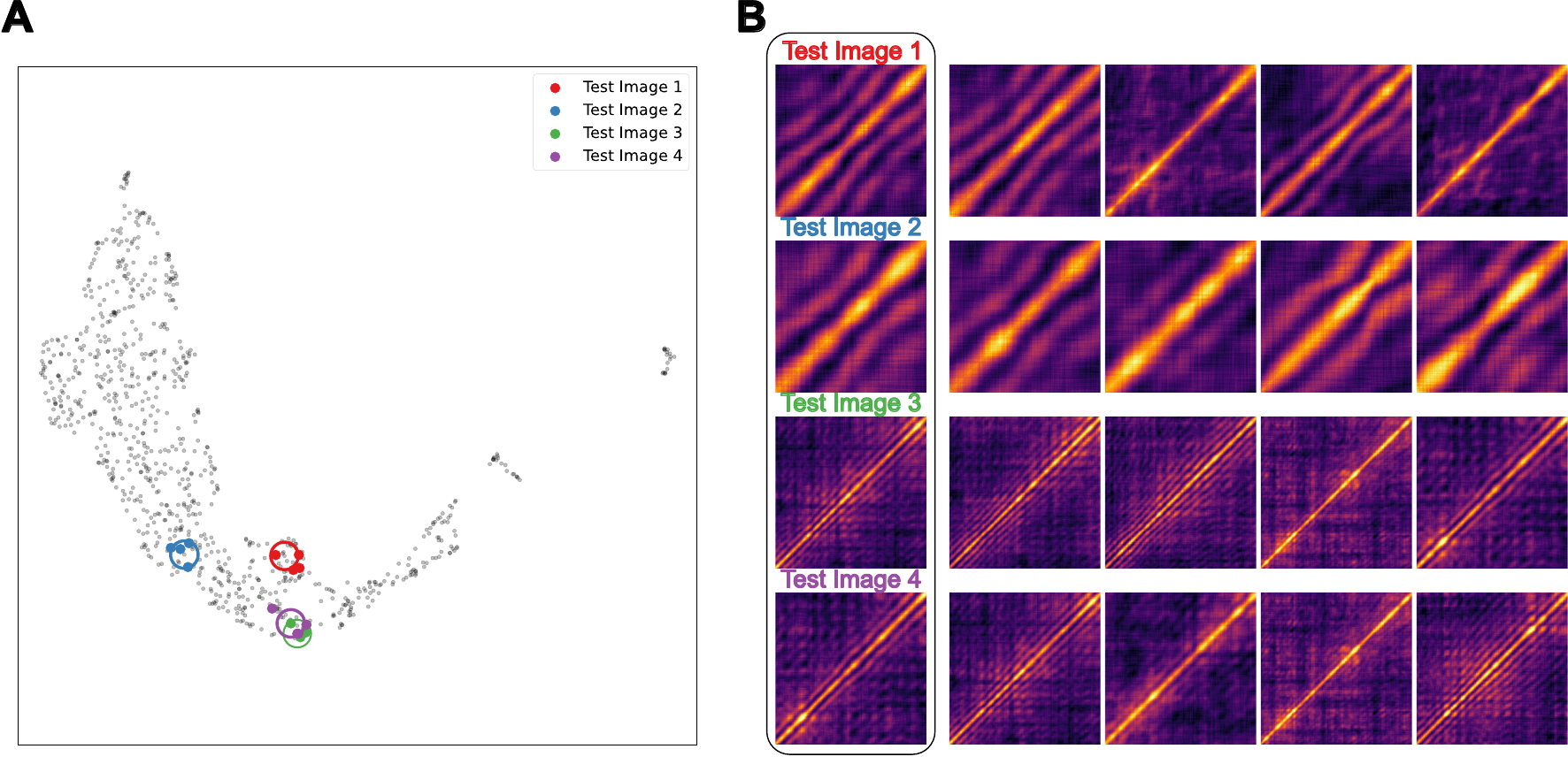}
    \caption{Distances in the latent space are used to suggest similar images to user-specific images of interest.  In A.), sample heterodyne $C_2$ are plotted in the UMAP visualization of the latent space as large open circles.  Nearest neighbors (calculated by euclidean distance in the latent space) are shown as correspond solid points.  In B.), sample images are displayed alongside their nearest neighbors to evaluate similarity.}
    \label{fig:similar_images}
\end{figure}

In addition to the difficulty of quantifying dynamics from XPCS data, another major bottleneck for the analysis of synchrotron scattering data is the amount of raw data which is collected, and then must be processed, reduced and analyzed.  For context, advanced x-ray detectors used at APS can collect up to 10-50 GB of raw scattering data per second, and many experiments may run continuously for hours.  With this in mind, we demonstrate how our unsupervised latent space analysis can be used to easily explore immense experimental datasets and identify relevant temporal features from time-resolved data.  Since the autoencoder learns to encode over-arching features of the entire data distribution, we can use the latent space distance between a user-specified sample image and other points as a metric of similarity to identify other experimental conditions which produce the same behavior.  Here we focus on the identification of $C_2$ showing heterodyning (see \textit{Test Images} in Figure \ref{fig:similar_images}B)\cite{livet_homodyne_2007}. Heterodyne XPCS features are created from the interference between the scattering signal of moving particles and the scattering signal from a stagnant reference.  In $C_2$ plots, it is typically manifested as fringes parallel to the diagonal observation time axis.  Test images were selected to have similar overall appearance (all show heterodyning), yet still represent significantly different behaviors: Images 1 and 2 show lower frequency fringes than Images 3 and 4, but the intensity along the diagonal band is unique in each test image.  If our model accurately encodes $C_2$, all four test images should appear close together in the latent space, with Images 1-2 and Images 3-4 being even closer together.  Figure \ref{fig:similar_images}A shows each test image (shown as large open circles) in the UMAP embedding of the latent representation, and the corresponding four nearest neighbors measured by euclidean distance in the latent space (the UMAP distribution is only used for visualization, all distances are calculated based on the AE encoding).  The apparent wide spread in neighborhoods for Test Images 1 and 2 can be attributed to the inability of the UMAP visualization to accurately represent the structure of the 64-dimensional latent space in a two-dimensional embedding.

XPCS datasets may contain thousands of unique $C_2$, making it infeasible to comb through all the data to identify instances with similar characteristics.  One can imagine an extension of our automated approach where the user defines a $C_2$ topology of interest and defines a distance from the sample $C_2$ to create a neighborhood of $C_2$ which are likely to look similar. After identifying closely-related $C_2$, corresponding measurements showing experimental time, shear stress, shear rate, viscosity, etc., could be aggregated.  Observing the distribution of system properties throughout the latent space will help scientists to link complex $C_2$ patterns to measurable changes in the rheology.

\section{\label{sec:Discussion}Discussion}

Upon first examining $C_2$, it is clear that even in complex non-equilibrium relaxation there are over-arching patterns that are similar between individual correlation functions (for example, see Test Images in Figure \ref{fig:similar_images}B).  With this in mind, our goal in clustering the data is to produce broad classes which can help us identify subsets of $C_2$; from here, more in-depth analysis can be performed, for example, to discover how experimental parameters, such as shear stress and viscosity vary within and between clusters.

While this type of analysis is helpful for making generalizations across the range of dynamics, we must interpret this statistical distribution of properties for each class in light of what the class represents; the k-means algorithm separates data by qualitative relaxation rate, but the boundaries between classes are difficult to correlate with physical features since they appear as essentially horizontal lines in the clustering visualization (Figure \ref{fig:initial_clustering})\cite{bin_waheed_machine_2019}.

The main challenge of understanding structural dynamics via XPCS usually lies in the interpretation of the correlation results. While $C_2$ from equilibrium dynamics depend only on the delay time between detector frames and can be reduced to 1D intensity autocorrelation function (also known as $g_2(\tau)$) to simplify the analysis, $C_2$ from \textit{in-situ} dynamics are usually far from equilibrium states where the same reduction is no longer applicable and methods for quantification of such dynamics are either unavailable, or difficult to interpret\cite{evenson_x-ray_2015,Ruta2020-oa,Zhang2017-oz,madsen_beyond_2010}. Several researchers have used XPCS data to suggest that glassy and jammed systems evolve through intermittent, or avalanche, dynamics wherein the system is nearly stationary until random structural fluctuations enable significant reconfiguration of the entire system\cite{song_microscopic_2022,evenson_x-ray_2015,cubuk_identifying_2015}.  By definition, these 'avalanches' are rare events, making them difficult to characterize.  Understanding rare events requires first understanding the mean behavior of the system, and then the ability to accurately detect and measure behavior outside the mean - this requires a rigorous statistical depiction of the system behavior based on a compilation of many unique measurements.
In our case, where the goal is to understand the link between stochastic, micro-scale structural changes and the evolution of macro-scale properties, characterization of intermittent dynamics becomes even more complicated, since bulk property measurements show contributions from the entire system, while XPCS experiments measure dynamics in only the small region of the sample illuminated by x-rays.  Considering that the measurable property is the average value of the measurement at each point in the system, it is unlikely that the intermittent dynamics identified in a single $C_2$ measurement will directly correspond with observable rheological changes measured across the whole system.  Similar to Figure \ref{fig:rheo_analysis}, where analysis of the distribution of classes allows us to understand the relationship between structure and dynamics, only through statistical analysis can the evolution of the entire system be inferred from an incomplete set of measurements.  While enough data exists to build these types of distributions, the time required for manual analysis limits the amount of data that can feasibly be considered.  Our method of automated $C_2$ processing and classification represents a first step towards statistical analysis of experimental data which will allow clear, quantitative understanding of relaxation dynamics and characterization of the mechanisms which lead to structural reconfiguration and subsequent changes in properties of the system.

\section{Conclusion}

We presented an unsupervised procedure for the automation of XPCS data interpretation.  The workflow allows us to explore the structure and distribution of large experimental datasets that would be difficult to otherwise interpret, and understand the dynamics of an evolving system.  As characterization instrumentation continues to improve, the amount of data collected in a single experiment will grow exponentially, yet the feasibility of manual analysis remains stagnant.  Therefore, automation of as much of the data analysis process as possible is imperative to fully utilize modern experimental equipment.  Our work using AI to guide the initial stages of data exploration and qualitative analysis represents an important step towards increasing the amount of available data which can actually be used, and presents a framework for parsing large datasets. As each $C_2$ dataset is associated with many metadata parameters (such as collection time, position in the sample, viscosity, shear stress, volumetric concentration, particle size, etc.), visualization of the latent space is key for the explanation of the relationships between parameters.  More importantly, this visualization and encoding framework is flexible and can be applied to experiments on other classes of materials, or even on different types of experimental data; while our analysis clearly shows how unsupervised deep learning can be used to extract rheological information from $C_2$, our method is a generic image processing framework which requires no physical information, and can therefore be applied to any experimental data which can be represented in two-dimensional/image space.

While our unsupervised clustering enables fast and reliable understanding of complex datasets and extraction of physical information on a qualitative level, future work must focus on bridging the gap between recognition of qualitative trends and defining quantitative models for complex processes.  Recent research has focused on the development of machine learning and statistical foundations for capturing, describing, and predicting non-linear dynamics\cite{brunton_discovering_2016,cranmer_discovering_2020}.  Moving forward, the combination of qualitative AI for guiding data analysis and quantitative AI for developing physical models of these systems will enable research that makes more complete use of acquired data leading to more detailed descriptions of material behavior.

\section{\label{sec:Methods}Methods}

\subsection{\label{sec:rheology}Rheology Experiments}
A sample of silica nanoparticles (200-300 nm) dispersed in polyethylene glycol (M.W. = 200) at volume fraction 60.5\% is used to study the dynamics of glassy systems.  The sample was loaded into a poly carbonate cylindrical Couette cell with a bob and cup (5.5 mm and 5.7 mm radii, respectively).  The shear cell was driven by an Anton Paar MCR 301 rheometer.  The x-ray beam is aligned at the center of the shear cell, so the detector plane is in the $q_{v}$ - $q_{\Delta x v}$ direction.  The sample is sheared under various conditions including preshear, steady shear ramp, and start-up shear.  After the shear sequence, the shear rate was set to zero, and the XPCS experiments were conducted to monitor the dynamics of particles at various positions of the sample.  Through XPCS measurements, the rheometer constantly monitors the stress relaxation process.

\subsection{\label{sec:xpcs_measurement}X-ray Photon Correlation Spectroscopy on Silica Nanoparticle Glass}

The XPCS measurement was performed at Beamline 8-ID-I of Advanced Photon Source, Argonne National Laboratory. An x-ray beam was generated by tandem 33 mm period, 2.4 m length undulators and was first deflected from a plane silicon mirror at an angle of 5 mrad then filtered  through a Ge(111) monochromator with a relative bandpass of 0.03\% to select a longitudinally coherent x-ray beam with a photon energy of 11 keV. The beam was then chopped horizontally to match the transverse coherence length at the entrance of the x-ray focusing optics (Beryllium Compound Refractive Lenses) and focused along the vertical direction, resulting in a 10 $\mu$m $\times$ 10 $\mu$m footprint on the sample with a total flux of $1.2\times10^{10}$ photons per second. 

The scattered x-ray intensities were collected at a distance of 8 m from the sample using a Lambda 750k photon-counting detector with 55 $\mu$m pixel size and 512$\times$1536 pixels \cite{Pennicard2013-px}. The XPCS analysis focuses on the region of detector pixels (Region of Interest, ROI) within the vicinity of the first peak in the structure factor (0.019 nm$^{-1} < Q < $ 0.029 nm$^{-1}$), and the ROI was further partitioned into 18 smaller ROIs in the angular direction (20$^{\circ}$ width) to account for the azimuthal asymmetry of the dynamics resulted from the rheological shear. $C_2$ is calculated from the multiplication of normalized intensity fluctuation $D(\vec{Q}, t)$ averaged over the entire ROI \cite{Zhang2017-oz,Fluerasu2005-ua}:

\begin{equation}
C_2(t_1, t_2) = \langle D(\vec{Q},t_1)\cdot D(\vec{Q},t_2) \rangle _{i,j}
\end{equation}

where $\langle ... \rangle _{i,j}$ indicate the pixel average. $D(\vec{Q}, t)$ is defined as:

\begin{equation}
D(\vec{Q},t) = \frac{I(\vec{Q},t)-\langle I(Q)\rangle_t }{\langle I(Q)\rangle_t }
\end{equation}

where $\langle I(Q)\rangle_t$ is the 1D Small-angle x-ray scattering (SAXS) intensity at the pixel with momentum transfer $\vec{Q}$, i.e., azimuthal average of the time-average from the detector frame sequence. 

\subsection{\label{sec:Dataset}Machine Learning Dataset Construction}
All $C_2$ in the dataset were measured on scattering patterns from silica sphere suspension at differing volume fractions and rheological conditions. 2000 unique $C_2$ were randomly selected from the entire volume of raw data.  Time correlations are calculated on groupings of 5000 frames at a time, so raw $C_2$ data is an image with 5000 $\times$
5000 pixels.  Two random crops of 256 $\times$ 256 pixels were taken along the diagonal from each raw $C_2$ to generate an initial dataset of 4000 256 $\times$ 256 $C_2$ images.  Severe data augmentation was required to capture the off-diagonal $C_2$ features present in real data.  Using the base set of 4000 training images, the model was not able to accurately represent off-diagonal features in $C_2$.  Augmenting the data by a factor of 10 (obtained by randomly shifting the data along the diagonal) showed minimal improvement, while augmentation by a factor of 100 (final training set containing 400,000 examples) produced accurate reconstructions after training.  Images were normalized such that the intensity distribution of each images ranged from 0 to 1.

\subsection{\label{sec:CNN_development}CNN Autoencoder Model}
We used a standard hourglass-style convolutional neural network as our autoencoder architecture.  This model uses four stages in both the encoding and decoding networks, where each stage consists of two convolutional layers.  In order to rapidly reduce the dimensionality of the data, and reduce the number of trainable parameters, we applied max-pooling to reduce the size of images by a factor of four after each stage; in the decoding model, it was found that upsampling after the convolutional layers performed better than using transpose convolution layers to upsample the images\cite{odena_deconvolution_2016,shi_is_nodate}.    Increasing either the number of convolutional layers or the number of filters per layer was found to degrade the quality of output image reconstructions; even with the augmented dataset model convergence was not stable as the size of the model increased.

We trained models with latent dimensions varying from 2 - 1024 (increasing in powers of two) to optimize the expressive power of the latent representation.  After training each model, mean squared error was evaluated on a test data set and the mean of the error was plotted as a function of latent dimension.  As shown in Supplemental Figure 2, the error rapidly decreased and leveled off at a latent dimension of 16.  We chose to use a bottleneck layer of size 64 for the final model to balance high accuracy with the complexity of the latent representation.  

The model was trained on the 100-times augmented dataset for 50 epochs using a cyclic learning rate in the Pytorch DL framework\cite{Paszke2019}.  Mean squared error loss was used to optimize the weights.  Specific hyperparameters can be found in the training script provided \textit{via} github.
All models were trained using a single NVidia A100 GPU using the Argonne National Laboratory LCRC cluster.

\subsection{Clustering and Classification}
After training the autoencoder, $C_2$ images were passed through the encoder stage only to produce the latent representation of the dataset. KMeans clustering was initially applied using the scikit-learn library with the number of clusters ranging from 2 to 12\cite{pedregosa_scikit-learn_2011}.  Plotting distortion as a function of number of clusters, the ideal number of clusters was determined to be in the range of four - six using the elbow method (Supplemental Figure 3).  Samples from each cluster were drawn to evaluate similarity within each cluster.  Uniform Manifold Approximation and Projection (UMAP) was used to project the 64-dimensional latent space into a two-dimensional visualization to inspect the quality of the clustering results and the position of optimized cluster centers.

\section*{Data Availability} 
Python scripts for reproducing analyses presented in tis paper are available at \url{https://github.com/jhorwath/XPCS_Clustering}.  Raw XPCS and rheology data is available at \url{https://anl.box.com/s/dhqahh467gnv0srz0tct1ymaofgr07te}

\section*{Acknowledgments}
This research used resources of the Advanced Photon Source, a U.S. Department of Energy (DOE) Office of Science user facility and is based on work supported by Laboratory Directed Research and Development (LDRD) funding from Argonne National Laboratory, provided by the Director, Office of Science, of the U.S. DOE under Contract No. DE-AC02-06CH11357. WC and HH were partially supported on XPCS data collection and analysis by the U.S. Department of Energy, Office of Science, Office of Basic Energy Sciences, Materials Science and Engineering Division.  We thank Nina Andrejevic and Saugat Kandel for helpful discussions on latent space analysis and CNN architecture optimization, respectively. 

\section*{Author contributions}
All authors contributed to conception of the research topic.  Neural network development and training was performed by J.P.H. and M.J.C., and clustering analysis was performed by J.P.H. with input from M.J.C., Q.Z., and S.N.  X-M.L., H.H., and S.N. collected experimental data.  All authors contributed to analysis of the results and preparation of the manuscript.

\section*{Competing interests} 
The authors declare that they have no competing financial interests.

\bibliographystyle{naturemag}
\bibliography{Library.bib,reference_2,cd_refs_2, scibib}

\end{document}